\documentclass[12pt]{iopart}

\usepackage{graphicx}  
\begin{document}

\title{100 Years of dimensional analysis: new steps toward empirical law deduction}

\author{M Taylor$^1$, A I D\'iaz$^2$\footnote{Sabbatical address: Institute of Astronomy, University of Cambridge, Madingley Road, Cambridge, CB3 0HA, UK}, L A J\'odar-S\'anchez$^3$ and R J Villanueva-Mic\'o$^3$}
\address{$^1$ Departamento de Astrof\'isica Molecular e Infrarroja (DAMIR), Instituto de Estructura de la Materia (IEM), Consejo Superior de Investigaciones Cient\'ificas (CSIC), Calle Serrano 121, Madrid 28006, Spain}
\address{$^2$ Grupo de Astrof\'isica, Departamento de F\'isica Te\'orica, Facultad de Ciencias, Universidad Aut\'onoma, Cantoblanco, 28049 Madrid, Spain}
\address{$^3$ Institute of Multi-disciplinary Mathematics, Polytechnic University of Valencia, Camino de Vera, Valencia 46022, Spain}
\ead{michael@damir.iem.csic.es}

\begin{abstract}
On the verge of the centenary of dimensional analysis (DA), we present a new matrix generalisation of the Buckingham Theorem. The proof is based on a solution we have found for inverting non-square block matrices and gives rise naturally to a new pair of transforms - the similarity transform (S) that converts physical dimensional data into dimensionless space and its inverse ($S^{'}$). Although it is well known that DA: a) reduces the number of free parameters, b) guarantees scale invariance through dimensional homogeneity and c) extracts functional information encoded in the dimensionless grouping of variables, scientists seem to be unaware that the scaling laws provided by DA are degenerate and therefore not unique. We demonstrate that the inverse transform $S^{'}$ is responsible for the non-uniqueness and we show how reference to observational data is sufficient to to break the degeneracy inherent in transforming back to dimensional (physical) space. As an example, we show how the underlying functional form of the Planck Radiation Law can be deduced in only a few lines using the matrix method and without appealing to first principles - thus demonstrating the possibility of a priori knowledge discovery. It is hoped that the proof presented here will provide new impetus to the pursuit of inverse problems in physics.
\end{abstract}

\vspace{2pc}
\maketitle

\section{Introduction}
The outstanding advances in experimental and theoretical physics in the second half of the $20^{th}$ Century have meant that we have been able to refine and improve the precision of the laws of physics (as evidenced by for example 15 decimal place accuracies for some atomic constants. However, the complexity of physical models has grown drastically at a rate which is proportional to the rate of accumulation of high resolution observational data meaning that exact, asymptotic and, in many cases, even approximate analytic expressions for the greater number of physical effects we are now able to include in our models, are difficult to deduce. As a result, many of us are resorting to systems-based modeling using for example holistic approaches such as complexity theory, or else are simulating interactions numerically using cellular automata, finite element analysis or Monte Carlo methods in an attempt to gain qualitative insight into complex natural phenomena. Furthermore, despite the giant technological leaps, observational errors, particularly in the field of astrophysics are of the order of 10\%, the empirical relations we are obtaining are still very approximate (order of magnitude) and in their infancy \cite{TAYLOR2007}. In budding areas of research, DA is often used to obtain functional dependencies between the physical variables and the resultant scaling relations are then used to estimate the value of a given parameter. Although often, "back-of-the-envelope" in nature, the approach helps to constrain the problem. In this paper, we generalize the Buckingham Theorem upon which DA is based and uncover an important problem at its crux - that of the non-uniqueness of scaling laws. Though this may appear to be catastrophic, we present a methodology for its resolution.\\

\section{Dimensional analysis (DA)}
DA appeared early in the thoughts of physicists \cite{FOURIER1822}, and since then has been studied and used by some of the most famous of them (for example, Maxwell \cite{MAXWELL1868}, Rayleigh \cite{RAYLEIGH1899} and Einstein \cite{EINSTEIN1911}). The first mathematical formulation of DA was published exactly a century ago \cite{MOROZOV1908} but it is often associated with the first general exposition of the ideas published in \textit{Nature} by Buckingham \cite{BUCKINGHAM1914} (see \cite{POBEDRYA2006} for a discussion of its origins). As such the fundamental theorem is often referred to as the Buckingham Theorem which has since been further developed in many textbooks \cite{GORTLER1975}, \cite{HUNTLEY1952}, \cite{SZIRTES1997}.\\
\newline\noindent
Although scaling relations obtained by DA are prevalent in physics and astrophysics, a citation search in NASA ADS of all physics, astrophysics and electronically archived papers, for example, shows that only 35 refereed articles have ever cited Buckingham's original paper and all have failed to address the problem of the uniqueness of dimensionless numbers. This  means that scaling relations used by or derived in these papers require great caution. The problem of uniqueness \textit{has} been partly addressed in the pioneering work of Stephan Rudolph and co-workers \cite{HERTKORN1998}, \cite{RUDOLPH1996}, \cite{RUDOLPH2000} from the engineering design community. In this paper, we show using a matrix generalisation of DA precisely where the degeracy comes from and how it may be addressed. Before doing so, we first clarify some of the lesser-known pitfalls associated with dimensionless numbers that, as we shall see, are the key to an efficient pre-processing of data.\\

\subsection{Dimensionless numbers}
A dimensionless number (or more precisely, a number with the dimensions of "1") has no physical units. It is typically defined as the ratio of quantities having units of identical dimension whose units then cancel. Whenever one measures a physical quantity it is with reference to a dimensional standard so that ultimately, we always work with dimensionless numbers and ratio scale, physical quantities. It is essential that the units are the same in both the numerator and denominator, such as m/m to avoid errors associated with unit conversions (e.g. cm/m and mm/cm are also dimensionless but introduce scaling factors of 1/100 and 1/10 respectively with respect to m/m). Already, here, we can see a degeneracy creeping in. We can, in fact, construct an infinite set of such length ratios that would all be dimensionless but which would differ by a constant of multiplication. Furthermore, a physical quantity that may be dimensionless in one system of units, may not be dimensionless in another system of units. For example, in the non-rationalized CGS system of units, the unit of electric charge (the Stratcoulomb) is defined in such a way so that the permittivity of free space $\epsilon_{0}=1/4\pi$ whereas in the rationalized SI system, $\epsilon_{0}=8.85419\times 10^{-12}F/m$. In systems of natural units (e.g. Planck units or atomic units), the physical units are defined in such a way that several fundamental constants are made dimensionless and set to "1" thus removing these scaling factors from equations. While this is convenient in some contexts, abolishing all or most units and dimensions makes practical physical calculations more error prone \cite{BAEZ2002}, especially, as we will see, when the goal is to pre-process data and extract empirical relations. The next section describes Buckingham's simple and elegant theorem concerning dimensionless numbers.\\

\subsection{The Buckingham $\pi$-Theorem}
In the physical sciences, the two sides of any equation must be commensurable or have the same dimensions (the principle of dimensional homogeneity) for all valid functional relationships \cite{BUCKINGHAM1915}. This led Buckingham to prove \cite{BUCKINGHAM1914} that, from the existence of a complete and dimensionally homogeneous function, $f$ of exactly $n$ physical variables $\{x_{n}\}\in\Re^{+}$,
\begin{equation}
f(x_{1},x_{2},\cdots,x_{n}) = 0,
\label{eqn 1}
\end{equation}
there exists a corresponding functional relationship $F$ of $m<n$ dimensionless numbers $\{\pi_{m}\}\in\Re^{+}$,
\begin{equation}
F(\pi_{1},\pi_{2},\cdots,\pi_{m}) = 0,
\label{eqn 2}
\end{equation}
where $m=n-r$ is reduced by the number $(r)$ of dimensionally independent variables in $\left\{x_{n}\right\}$. The restriction to positive values of the variables can be satisfied universally by suitable coordinate transformations. The Buckingham Theorem holds for all dimensionally correct equations of physics \cite{PAWLAWSKI1971}, \cite{GORTLER1975}. The $\pi$-groups are dimensionless numbers or "similarity numbers" which are the scaling invariants of a physical problem. Although simple, in essence, Buckingham's Theorem is extremely profound. From our observations of reality we categorise data into homogeneous sets that we identify with different physical variables $\{x_{n}\}$. By constructing a qualitative model relating the variables, we find $p$ different various physical properties $f(x_{1},\cdots,x_{n})_p$. In the context of a chosen system of units (e.g. SI) and frame of reference (e.g. coordinate system or inertial frame), we are able to offer a quantitative description of reality $f(x_{1},\cdots,x_{n})=0$. Finally, in order to correctly evaluate reality on all scales, we must find the mimimal and invariant description in terms of dimensionless numbers $F(\pi_{1},\cdots,\pi_{m}) = 0$. It is at this point that we see the importance of DA - only once the function $F$ is found have we successfully decoded the empirical law describing the data.\\

\subsection{A Matrix Generalization of DA}
\begin{figure*}
\centering
\vspace{5cm}
\includegraphics{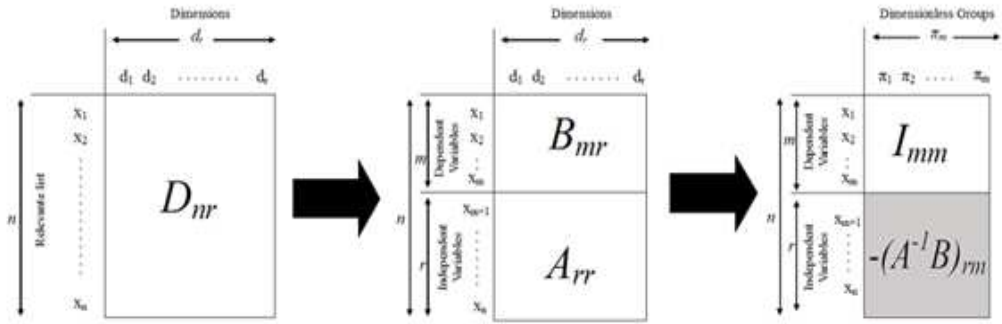}
\caption{Construction of the dimensionless matrix from the dimensional one using rank-preseving operations}\protect\label{dimension_matrix}
\end{figure*}

DA provides a procedure to generate dimensionless numbers from a list of relevant physical variables together with their respective dimensions. Several authors have heuristically arrived at matrix expressions to calculate dimensionless numbers \cite{SZIRTES1997},\cite{RUDOLPH1996} but a full mathematical proof has, until now, been lacking. The reason appears to be two-fold: 
\begin{enumerate}
	\item for simple problems, a dimensionless combination of variables can often be found by inspection (e.g. a ratio of lengths),
  \item a generalised matrix inverse for the resultant non-square, degenerate matrices involved (see below) has, until now, only been calculable numerically and therefore approximately. 
\end{enumerate}
It is possible that the failure of many authors to address the problem of the non-uniqueness of dimensionless numbers may be traced back to the absence of such a proof, which we present below. In what follows, we use the notation employed in \cite{SZIRTES1997}.\\
\newline
The dimensional equations of the physical variables in equation 1 (with "[ ]" representing "dimensions of") can be expressed as:
\begin{eqnarray*}
\left[x_{1}\right]&=&d_{1}^{\alpha_{11}}d_{2}^{\alpha_{12}}...d_{r}^{\alpha_{1r}}\\
\left[x_{2}\right]&=&d_{1}^{\alpha_{21}}d_{2}^{\alpha_{22}}...d_{r}^{\alpha_{2r}}\\
&\vdots&\\
\left[x_{n}\right]&=&d_{1}^{\alpha_{n1}}d_{2}^{\alpha_{n2}}...d_{r}^{\alpha_{nr}},
\label{eqn 3}
\end{eqnarray*}
\noindent
where the $\{d_{j},j=1,\ldots,r\}$ are the base dimensions having exponents $\alpha_{ij}$. In the SI system of units, there are 7 distinct base dimensions: mass $M[Kg]$, length $L[m]$, time $T[s]$, temperature $\theta[K]$, electrical current $I[A]$, concentration $N[mol]$ and light intensity $J[cd]$. So, for example, a pressure $P$ which is measured in $N/m^{2}\equiv[kg^{1}m^{-1}s^{-2}]$ would have a dimensional equation of the form $[P]=M^{1}L^{-1}T^{-2}\theta^{0}I^{0}mol^{0}J^{0}$. In this case, the exponents $\alpha_{ij}$ of the dimensions of pressure are given by the row vector $\left(1,-1,-2,0,0,0,0\right)$. The matrix formed with the exponents $\alpha_{ij}$ of the dimensions $d_{j}$ of all physical variables $x_{i}$ relevant to a problem, we call the dimension matrix $\tilde{D}$:
\begin{equation}
    \tilde{D}= \left [ \begin{array}{cccc}
    \alpha_{11}&\alpha_{12}&\ldots&\alpha_{1r}\\
    \alpha_{21}&\alpha_{22}&\ldots&\alpha_{2r}\\
    \vdots&\vdots&\ldots&\vdots\\
    \alpha_{n1}&\alpha_{n2}&\ldots&\alpha_{nr} \end{array} \right ].
\label{eqn 5}
\end{equation}
\noindent
$\tilde{D}$ has $n$ rows for the number of physical variables and $r$ columns for the distinct base dimensions. To satisfy dimensional homogeneity, we need to find the exponents $\epsilon_{i}$ that solve the equation,
\[
[x_{1}^{\epsilon_{1}}x_{2}^{\epsilon_{2}}...x_{n}^{\epsilon_{n}}]=
d_{1}^{q_{1}}d_{2}^{q_{2}}...d_{r}^{q_{r}},
\]
with $\{q_{j},j=1,\cdots,r\}$ being the sought combinations of fundamental dimensions (to be set equal to zero for dimensionless numbers). The condition, $q_{1}=q_{2}\cdots q_{j}=0$ then leads directly to dimensional homogeneity:
\begin{equation}
	[x_{1}^{\epsilon_{1}}x_{2}^{\epsilon_{2}}...x_{n}^{\epsilon_{n}}]=1.
\label{homogeneity}
\end{equation}
A dimensionless (dimension "1") $\pi$ group is therefore given by,
\begin{equation}
	\pi=x_{1}^{\epsilon_{1}}x_{2}^{\epsilon_{2}}\cdots x_{n}^{\epsilon_{n}}.
\label{pi_group}
\end{equation}
Our task here then is to find the exponents $\epsilon_{j}$. We proceed as follows. The dimensional equations for the physical variables $\left[x_{i}\right]$, when raised to the power $\epsilon_{i}$ become,
\begin{eqnarray*}
\left[x_{1}^{\epsilon_{1}}\right]&=&\left(d_{1}^{\alpha_{11}}d_{2}^{\alpha_{12}}...d_{r}^{\alpha_{1r}}\right)^{\epsilon_{1}}\\
\left[x_{2}^{\epsilon_{2}}\right]&=&\left(d_{1}^{\alpha_{21}}d_{2}^{\alpha_{22}}...d_{r}^{\alpha_{2r}}\right)^{\epsilon_{2}}\\
&\vdots&\\
\left[x_{n}^{\epsilon_{n}}\right]&=&\left(d_{1}^{\alpha_{n1}}d_{2}^{\alpha_{n2}}...d_{r}^{\alpha_{nr}}\right)^{\epsilon_{n}}.
\end{eqnarray*}
Evaluating the product $\prod_{i=1}^{n}\left[x_{i}^{\epsilon_{i}}\right]$ and grouping the base dimensions $d_{j}$ we obtain,
\begin{eqnarray*}
    [x_{1}^{\epsilon_{1}}x_{2}^{\epsilon_{2}}...x_{n}^{\epsilon_{n}}]&=& d_{1}^{\alpha_{11}\epsilon_{1}+\alpha_{21}\epsilon_{2}+\ldots+\alpha_{n1}\epsilon_{n}}\\
    &\times&d_{2}^{\alpha_{12}\epsilon_{1}+\alpha_{22}\epsilon_{2}+\ldots+\alpha_{n2}\epsilon_{n}}\\
    &\vdots&\\
    &\times&d_{r}^{\alpha_{1r}\epsilon_{1}+\alpha_{2r}\epsilon_{2}+\ldots+\alpha_{nr}\epsilon_{n}}.
\end{eqnarray*}
Associating $q_{j}$ with the exponents of the base dimensions $d_{j}$ then we arrive at the following linear system of equations,
\begin{eqnarray*}
	\alpha_{11}\epsilon_{1}+\alpha_{21}\epsilon_{2}+\ldots+\alpha_{n1}\epsilon_{n}&=&q_{1}\\
	\alpha_{12}\epsilon_{1}+\alpha_{22}\epsilon_{2}+\ldots+\alpha_{n2}\epsilon_{n}&=&q_{2}\\
	&\vdots&\\
	\alpha_{1r}\epsilon_{1}+\alpha_{2r}\epsilon_{2}+\ldots+\alpha_{nr}\epsilon_{n}&=&q_{r},
\end{eqnarray*}
which in matrix form are given by,
\begin{equation}
	\tilde{D}^{T}\vec{\epsilon}=\vec{q}.
\label{eqn 7}
\end{equation}
The $\epsilon_{j}$ can then be obtained from the inverse dimension matrix,
\begin{equation}
	\vec{\epsilon}=\left(\tilde{D}^{T}\right)^{-1}\vec{q}.
\label{eqn 8}
\end{equation}
provided that $\left|\tilde{D}^{T}\right|\neq 0$. The vector $\vec{\epsilon}$ containing the $\epsilon_{j}$ then allows identification of a single dimensionless number. However, we know from Buckingham's Theorem that there are exactly $\textit{m}$ such groups. We therefore require a solution vector for each of the $\pi$-groups. Thus, in equation (\ref{eqn 8}), the vectors $\vec{\epsilon}$ and $\vec{q}$ must become matrices $\tilde{\epsilon}$ and $\tilde{q}$ of dimension ($n\times m$) and ($r\times m$) respectively. The result is that dimensional homogeneity will then lead to the set of ($j=1,\cdots,m$) dimensionless groups $\pi_{j}$ given by,
\begin{equation}
	\pi_{j}=x_{1}^{\epsilon_{1j}}x_{2}^{\epsilon_{2j}}\cdots x_{n}^{\epsilon_{nj}}.
\label{homogeneity2}
\end{equation}
Our next task is to invert the dimension matrix $\tilde{D}^{T}$. However, it is non-square and \textit{degenerate} since the number of physical variables is always greater than the number of related base dimensions \cite{BUCKINGHAM1915}. The generalised inverse of a non-square matrix is far from trivial to find and, until now, only numerical approximations have been available \cite{MITRA1974}. However, for a special class of inverses \cite{CAMPBELL1991} we know that the matrix equation $\tilde{A}\tilde{x}=\tilde{b}$ for a non-square ($n\times m$) matrix $\tilde{A}$, has a solution \textit{iff} $\tilde{A}\tilde{A}^{+}\tilde{b}=\tilde{b}$ where the $(m\times n)$ matrix $\tilde{A}^{+}$ is the Moore-Penrose inverse of $\tilde{A}$ such that $\tilde{A}\tilde{A}^{+}\tilde{A}=\tilde{A}$. In a second remarkable paper published in the same year \cite{JODAR1991}, it was shown that the Moore-Penrose inverse $\tilde{A}^{+}$ is rarely equal to the generalised matrix inverse $\tilde{A}^{-1}$. The matrix solution provided by \cite{SZIRTES1997} obtains the Moore-Penrose inverse. Although $\tilde{A}^{+}$ can always be found numerically by SVU decomposition for well-defined problems, our interest here is the closed (exact) algebraic form.\\
\newline\noindent
Rather than seeking the Moore-Penrose inverse and then attempting to show its generality for the infinite set of solutions, we have been able to find a closed form for the generalized solution directly. The proof is short and proceeds as follows. We know from Buckingham's Theorem and the principle of dimensional homogeneity that each dimensionless $\pi$-group is constructed from an independent physical variable multiplied by a suitable combination of the dependent physical variables \cite{BUCKINGHAM1914}. We start then, by partitioning the dimension matrix $\tilde{D}^{T}$ into two inner block matrices: $\tilde{A}^{T}$ for the dependent variables and $\tilde{B}^{T}$ for the independent variables,
\begin{equation}
	\tilde{D}^{T}=\left[\begin{array}{cc} \tilde{B}^{T} & \tilde{A}^{T}\end{array} \right].
\label{eqn 10}
\end{equation}
The rank $r$ of the matrix $\tilde{A}^{T}$ is equal to the number of base dimensions $d_{j}$ involved in the problem. It is known \cite{MITRA1974} that if $\tilde{X}=\tilde{A}^{-1}$ is any matrix satisfying,
\begin{equation}
	\tilde{A}\tilde{X}\tilde{A}=\tilde{A},
\label{eqn 10a}
\end{equation}
then the linear system 
$\tilde{A}\tilde{x}=\tilde{b}$ has a solution \textit{iff}, $\tilde{A}\tilde{X}\tilde{b}=\tilde{b}$. Such as system has a general solution,
\begin{equation}
	\tilde{x}=\tilde{X}\tilde{b}+\left(\tilde{I}-\tilde{X}\tilde{A}\right)\tilde{y},
\label{eqn 10b}
\end{equation}
with $\tilde{y}$ arbitrary. In order to calculate the complete set of $\pi$ groups, we need to solve the matrix system equivalent of equation (\ref{eqn 7}),
\begin{equation}
	\tilde{D}^{T}\tilde{\epsilon}=\tilde{q}
\label{eqn 7b}
\end{equation}
such that,
\begin{equation}
	\tilde{\epsilon}=\left(\tilde{D}^{T}\right)^{-1}\tilde{q}.
\label{eqn 7c}
\end{equation}
Here, $\tilde{\epsilon}$ is a matrix of size ($n\times m$), $\tilde{q}$ is a matrix of size ($r\times m$) and $\tilde{D^{T}}$ is a block matrix of size ($r\times n$) partitioned into the ($r\times m$) matrix $\tilde{B^{T}}$ and the ($r\times r$) invertible matrix $\tilde{A^{T}}$. Taking into account that from equation (\ref{eqn 10a}),
\[\left[\begin{array}{cc} \tilde{B^{T}} & \tilde{A^{T}}\end{array}\right] \left[ 
\begin{array}{c}
0 \\ 
\tilde{A^{T}}^{-1}
\end{array}
\right] \left[\begin{array}{cc}\tilde{B^{T}} & \tilde{A^{T}}\end{array}\right] =\left[ \tilde{B^{T}},\tilde{A^{T}}\right]
\]
and that,
\[
\left[\begin{array}{cc} \tilde{B^{T}} & \tilde{A^{T}}\end{array}\right] \left[ 
\begin{array}{c}
0 \\ 
\tilde{A^{T}}^{-1}
\end{array}
\right] \tilde{q}=\tilde{I}\tilde{q}=\tilde{q},
\]%
then, (\ref{eqn 10b}) provides the general solution for equation (\ref{eqn 7c}):
\begin{eqnarray}
\tilde{\epsilon}  &=&\left[ 
\begin{array}{c}
0 \\ 
\tilde{A^{T}}^{-1}
\end{array}
\right] \tilde{q}+\left( \tilde{I}-\left[ 
\begin{array}{c}
0 \\ 
\tilde{A^{T}}^{-1}
\end{array}
\right] \left[ \tilde{B^{T}} \tilde{A^{T}}\right] \right) \tilde{y}  \nonumber \\
&=&\left[ 
\begin{array}{c}
0 \\ 
\tilde{A^{T}}^{-1}\tilde{q}
\end{array}
\right] +\left( \left[ 
\begin{array}{cc}
\tilde{I} & 0 \\ 
0 & \tilde{I}
\end{array}
\right] -\left[ 
\begin{array}{cc}
0 & 0 \\ 
\tilde{A^{T}}^{-1}\tilde{B^{T}} & \tilde{I}%
\end{array}
\right] \right) \left[ 
\begin{array}{c}
\tilde{y_{1}} \\ 
\tilde{y_{2}}
\end{array}
\right]   \nonumber \\
&=&\left[ 
\begin{array}{c}
0 \\ 
\tilde{A^{T}}^{-1}\tilde{q}
\end{array}
\right] +\left[ 
\begin{array}{cc}
\tilde{I} & 0 \\ 
-\tilde{A^{T}}^{-1}\tilde{B^{T}} & 0
\end{array}
\right] \left[ 
\begin{array}{c}
\tilde{y_{1}} \\ 
\tilde{y_{2}}
\end{array}
\right]   \nonumber \\
&=&\left[ 
\begin{array}{c}
\tilde{y_{1}} \\ 
\tilde{A^{T}}^{-1}\tilde{q}-\tilde{A^{T}}^{-1}\tilde{B^{T}}\tilde{y_{1}}
\end{array}
\right]   \nonumber \\
\tilde{\epsilon}&=&\left[ 
\begin{array}{c}
\tilde{y_{1}} \\ 
\tilde{A^{T}}^{-1}\left( \tilde{q}-\tilde{B^{T}}\tilde{y_{1}}\right) 
\end{array}
\right].
\label{eqn 10c}
\end{eqnarray}
Here, $\tilde{y_{1}}$ is an arbitrary matrix of size ($m\times m$). Equation ($\ref{eqn 10c}$) gives the whole set of
solutions of the matrix system ($\ref{eqn 7b}$). Here, we are concerned with obtaining the dimensionless numbers $\pi_{j}$ for which we need to set $\tilde{q}=0$. Inserting this into equation (\ref{eqn 10c}) we see that, in order to satisfy equation (\ref{eqn 7b}), the arbitrary matrix $\tilde{y_{1}}$ must be equal to the identity matrix $\tilde{I}$ since only then do we find that
\begin{eqnarray}
	\tilde{D}^{T}\tilde{\epsilon}&=&
	\left[\begin{array}{cc} \tilde{B}^{T} & \tilde{A}^{T}\end{array} \right]\left[ 
	\begin{array}{c}
	\tilde{I} \\ 
	-\tilde{A^{T}}^{-1}\tilde{B^{T}}
	\end{array}\right]\nonumber\\
	&=&\tilde{B}^{T}\tilde{I}-\tilde{A}^{T}\left(\tilde{A}^{T}\right)^{-1}\tilde{B}^{T}\nonumber\\
	&=&\tilde{B}^{T}\tilde{I}-\tilde{I}\tilde{B}^{T}=\tilde{q}=0.\nonumber
\end{eqnarray}
The condition $\tilde{q}=0$ is a special case for which the arbitrary matrix $\tilde{y_{1}}$ becomes specfic rather than arbitrary. This is exactly the case we are seeking to determine dimensionless combinations of the physical variables. The solution to our problem then, is given by,
\begin{equation}
	\tilde{\epsilon}(\tilde{q}=0)= \left[ 
	\begin{array}{c}
	\tilde{I} \\ 
	-\tilde{A^{T}}^{-1}\tilde{B^{T}}
	\end{array}
	\right].
\label{eqn 10d}
\end{equation}
The solution matrix $\tilde{\epsilon}$ has $n$ rows, one for each of the physical variables $x_{i}$ and $m$ columns, one for each of the dimensionless numbers $\pi_{j}$. The identity matrix $\tilde{I}$ in the upper half of $\tilde{\epsilon}$ guarantees that only one independent variable at a time is combined with dependent variables in constructing the $\pi$-groups. Furthermore, as we might expect, the solution matrix $\tilde{\epsilon}$ is a block matrix having an upper-diagonal form. In the case of the inversion by Guassian-Jordan elimination of a non-singular \textit{square} matrix $\tilde{A}^{S}$, then problem matrix is augmented with the identity matrix and inversion proceeds as follows:
\begin{eqnarray}
	\tilde{A}^{S}\rightarrow
	\left[\begin{array}{c|c} \tilde{A}^{S} & \tilde{I}\end{array}\right]\rightarrow
	\left[\begin{array}{c|c} \tilde{I} & (\tilde{A}^{S})^{-1}\end{array}\right].\nonumber
\end{eqnarray}
For our \textit{non-square} partitioned matrix $\tilde{D}^{T}$ then, in obtaining the inverse matrix, only the symmetric part of $\tilde{B}$ is effectively augmented by the identity matrix; the semantic structure of Guassian Elimination is maintained. Figure \ref{dimension_matrix} shows the overall process. The dimension matrix $\tilde{D}$, (left), is formed from the exponents $\alpha_{ij}$ of the base dimensions $d_{j}$ of the relevant physical variables $x_{i}$; one row for each variable and one column for each base dimension. The rank $r$ of the dimension matrix is equal to the number of base dimensions involved and is used to partition $\tilde{D}$ (centre) with the block matrix $\tilde{B}$ containing the exponents of the independent variables and the block matrix $\tilde{A}$ containing the exponents of the dependent variables. The solution matrix $\tilde{\epsilon}$  is then obtained and comprises an upper square identity matrix $\tilde{I}$ and a lower sub-matrix $-(\tilde{A}^{T})^{-1}\tilde{B}^{T}$.\\
\newline\noindent
The various matrices used in the proof include the square matrices $I_{mm}$, $A_{rr}$ and the non-square matrices: $D_{nr}$, $D^´{T}_{nr}$, $(D^{T})^{-1}_{nr}$, $\epsilon_{nm}$, $q_{rm}$, $B_{mr}$ and $(\tilde{A}^{T})^{-1}\tilde{B}^{T}_{rm}$. The mathematical complexity of the problem of evaluating the dimensionless $\pi$-groups therefore lies in the calculation of the inverse matrix $(\tilde{A}^{T})^{-1}$ of dimension equal to the rank $r$ of $\tilde{D}$. The condition for $(\tilde{A}^{T})^{-1}$ to exist is that the determinant $\left|\tilde{A}^{T}\right|\neq 0$. This mathematical condition corresponds to the physical condition that we have correctly identified in $x_{n}$ the complete list of relevant physical variables required to satisfy dimensional homogeneity. Had we missed out a crucial variable, then the determinant would equal zero producing a singular and undefined inverse. We thus, have an additional check on the correct construction of the problem in dimensional space. Since the dimensionless groups form a reduced set (there being $m=n-r$ of them), they minimise the number of free parameters. The $\pi$-groups then provide the most compact formulation possible for the evaluation of physical laws. In the next section we show how the matrix method gives rise naturally to a pair of transforms: the Similarity Transform $S$ and its inverse $S^{-1}$ that allow us to reveal the origin of the problem of non-uniqueness associated with scaling laws.\\

\subsection{Similarity Theory and the Uniqueness Problem}
\begin{figure*}
\centering
\vspace{7cm}
\includegraphics{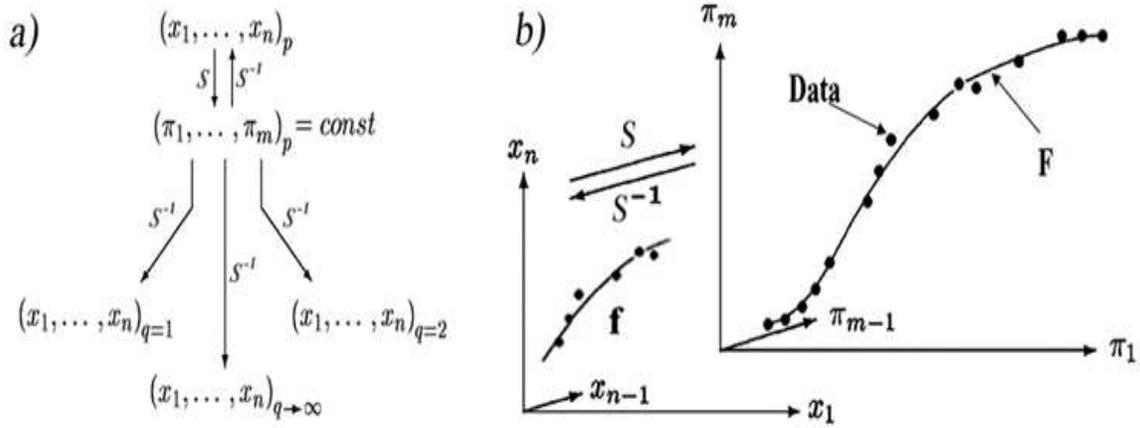}
\caption{Similarity transforms: dimensionality expansion $\pi^{-1}$ of case $p$ onto all completely similar cases $q=1,\cdots,\infty$: adapted from \cite{RUDOLPH2000}}
\label{uniq}
\end{figure*}

\noindent
We have described how functions of the type $f\left(x_{1},\cdots,x_{n}\right)=0$ or equivalently, $x_{n}=f\left(x_{1},\cdots,x_{n-1}\right)$, serve as quantitative models. The philosophical principle of causality \cite{BUNGE1957} is guaranteed when, for a complete list of physical variables $x_{n}$, the presence of cause $\left(x_{1},\cdots,x_{n-1}\right)$ \textit{always} results in effect $x_{n}$. Any question about the behavior of the physical object can be answered by the rule $f$, otherwise $f\left(x_{1},\cdots,x_{n}\right)=0$ would not represent the complete and correct physical model. This means that $F\left(\pi_{1},\cdots,\pi_{m}\right)=0$ represents the similarity transform of $f$ under which the physical content of the rule $F$ remains invariant \cite{RUDOLPH2000}. The similarity transform $S:X\mapsto\Pi$ from the space $X$ of physical variables $\{x_{i}\}$ to the dimensionless space $\Pi$ of dimensionless variables $\{\pi_{j}\}$ from equation (\ref{homogeneity2}) is given by,
\begin{equation}
	S:\pi_{j} = \prod^{n}_{i=1}x^{\epsilon_{ij}}_{i}
\label{eqn 16}
\end{equation}
\noindent
with $j\in\left[1,m\right]$. $S$ represents the transform of a space $\Re^{n}$ onto a space $\Re^{m}$ where $m=n-r$ corresponds to a dimensionality reduction leading to different points in $X$ being mapped to the same point in $\Pi$. From the Buckingham Theorem for the functional relationship between $\pi$ groups, \[F(\pi_{1},\cdots,\pi_{m}) = 0,\] any single $\pi$ group can. analogously, be expressed as a \textit{causal} function $F$ of the remaining groups:
\begin{equation}
\pi_{j}=F(\pi_{k\neq j})
\label{pi_other}
\end{equation}
with $k\leq m$. The $\pi$ groups are calculated from the physical variables $x_n$ using the Similarity Transform $S$ (equation \ref{eqn 16}). Since each $\pi$ group contains one and only one independent variable, we can find the inverse transform using their negative exponents in the product functio such that the other independent variables cancel out in the $j^{th}$ group,
\[
	x_{j}=\pi_{j}\prod^{r}_{i=1}x^{-\epsilon_{(i+m)j}}_{i+m}.
\]
Then, substituing for $\pi_{j}$ from equation (\ref{pi_other}), the inverse similarity transform $S^{-1}$ is given by,
\begin{equation}
	S^{-1}:x_{j}=\prod^{r}_{i=1}x^{-\epsilon_{(i+m)j}}_{i+m}F\left(\pi_{k}=\prod^{n}_{i=1}x^{\epsilon_{il}}_{i}\right)_{k\neq j}.
\label{eqn 16b}
\end{equation}
This then gives the scaling laws for the independent variables $x_{j}$ expressed as a function $F$ of the dependent variables combined in $\pi$ groups.\\
\newline\noindent
The inverse similarity transform $S^{-1}:\Pi\mapsto X$ of the dimensionless space $\Pi$ to the dimensional space of physical variables $X$ is not unique since it maps a space $\Re^{m}$ onto a larger space $\Re^{n}$; there being only $m$ dependent variables. A single point $p$ in $X$-space with $\left(x_{1},\cdots,x_{n}\right)_{p}$ is mapped via the $S$ transform onto its corresponding point $\left(\pi_{1},\cdots,\pi_{m}\right)_{p}$ in dimensionless space $\Pi$. The inverse transform $S^{-1}$ leads to the whole set of all completely similar points $\left(x_{1},\cdots,x_{n}\right)_{q=1,\cdots,\infty}$ in $X$ defined by the similarity condition $\left(\pi_{1},\cdots,\pi_{m}\right)_{p=constant}$. The function $F$ in $\Pi$-space is simpler than its counterpart $f$ in dimensional space $X$; being of lower dimensionality. Mathematically, the $\pi_{j}$=constant define a hypersurface in $X$ that contain all completely similar points and reflect the fact that there exists an infinite set of dimensionless numbers for a given physical problem \cite{RUDOLPH2000}. This can be seen by taking, for example, the dimensionless ratio $\pi_{L}=l_{1}/l_{2}$ of two lengths $l_{1}$ and $l_{2}$. When $l_{1}=3m$ and $l_{2}=4m$ or when $l_{1}=6m$ and $l_{2}=8m$, then in both cases $\pi_{L}=0.75$. It is the non-uniquness of the inverse mapping $\Pi\mapsto X$ that is responsible for the non-uniqueness of $F$ and hence of scaling laws deduced from the inverse transform $S^{-1}$. Figure \ref{uniq} shows a visual representation of how the asymmetry between $f$ and $F$ results in non-uniqueness of the dimensionless groups. It is this non-uniqueness that propagates back to scaling laws and the identification of empirical laws.\\
\newline\noindent
DA provides the mechanism for mapping a problem into a space where dimensional reduction can be exploited. Proper identification of the function $F$ then allows for a full problem evaluation obtained by inverse transformation back to dimensional space. Fourier transforms, Laplace transforms and the like all take advantage of the algebraic simplicity in the transformed space before inverse-transforming back to the problem space. However, it is the ambiguity that arises from the inverse transform that is ultimately responsible for the problem of non-uniqueness when the dependent variables are expressed in terms of the $\pi$ groups. As an illustration, in the next section we apply DA to the problem of Planck Radiation to show a) the ease of application of our matrix method, b) how the problem of degeneracy manifests itself,  c) how, despite this \textit{a priori} knwoledge discovery is still possible, and d) how reference to data can break the degeneracy and provide unique scaling laws.\\

\section{The Planck Radiation Law}
For Planck radiation, we know that the spectral intensity $u$ is related to the electromagnetic radiation emitted at different frequencies $\nu=c/ \lambda$ from a black body at a temperature $T$. Since the radiation is generated by oscillators having energy $k_{B}T$ per degree of freedom and is carried by photons travelling at the speed of light $c$ in packets of energy  $h\nu$, we expect that the total integrated spectral intensity will be a function of the following list of physical variables:
\begin{equation}
f(u,\lambda,h,c,k_{B},T)=0
\label{eqn 18}
\end{equation}
The dimensional equations of the physical variables are:
\begin{eqnarray}
\left[u\right]& = & M^{1} L^{-2} T^{-2} \theta^{0} = M L^{-2} T^{-2}\nonumber\\
\left[\lambda\right] & = & M^{0} L^{1} T^{0} \theta^{0} = L\nonumber\\
\left[h\right]& = & M^{1} L^{2} T^{-1} \theta^{0} = L = M L^{2} T^{-1}\nonumber\\
\left[c\right]& = & M^{0} L^{1} T^{-1} \theta^{0} = L T^{-1}\nonumber\\
\left[k_{B}\right] & = & M^{1} L^{2} T^{-2} \theta^{-1}\nonumber\\
\left[T\right] & = & M^{0} L^{0} T^{0} \theta^{1} = \theta,
\label{eqn 18b}
\end{eqnarray}
and give rise to the dimension matrix,
\begin{equation}
\tilde{D} = \begin{tabular}{ l | c c c c }
    & $M$ & $L$ & $T$ & $\theta$\\
  \hline
  $u$  & 1 & -2 & -2 & 0\\
  $\lambda$  & 0 & 1 & 0 & 0\\
  $h$  & 1 & 2 & -1 & 0 \\
  $c$ & 0 & 1 & -1 & 0 \\
  $k_{B}$ & 1 & 2 & -2 & -1 \\
  $T$ & 0 & 0 & 0 & 1 \\ 
\end{tabular}
\label{eqn 19}
\end{equation}
The next step involves seperation of the elements of $\tilde{D}^{T}$ into two block matrices: $\tilde{A}^{T}$ for the dependent variables and $\tilde{B}^{T}$ for the independent variables. Since there are $n=6$ physical variables in equation (\ref{eqn 18}) and $r=4$ base dimensions, the number of $\pi$ groups will be $m=n-r=2$. There are therefore $r=4$ dependent variables and $m=2$ independent variables. We are looking for an equation for the spectral intensity $u$ and so this will be one of the two independent variables; the other is the wavelength $\lambda$. As a check on the completeness of the list of physical variables, we calculate the determinant $|\tilde{A}^{T}|=1\neq 0$ as required for dimensional homogeneity and for the invertibility of $\tilde{A}$. We can now partition the dimension matrix $\tilde{D}$ and take the transpose,
\begin{eqnarray}
    \tilde{D}^{T} =\left[\begin{array}{c c} \tilde{B}^{T} & \tilde{A}^{T}\end{array} \right]
    =\left[\begin{array}{cc|cccc}
1 & 0 & 1 & 1 & 1 & 0\\
-2 & 1 & 2 & 1 & 2 & 0\\
-2 & 0 & -1 & -1 & -2 & 0\\
0 & 0 & 0 & 0 & -1 & 1\\
\end{array}
\right]
\label{eqn 20}
\end{eqnarray}
The matrix product $-(\tilde{A}^{T})^{-1}\tilde{B}^{T}$ then gives the solution matrix,
\begin{equation}
\tilde{\epsilon} = \left[\begin{tabular}{c}
  $\tilde{I}$\\
  $-(\tilde{A}^{T})^{-1}\tilde{B}^{T}$\\
\end{tabular}\right] = 
\begin{tabular}{ l | cc}
  & $\pi_{1}$ & $\pi_{2}$ \\
  \hline
  $u$  & 1 & 0 \\
  $\lambda$  & 0 & 1 \\
\hline
  $h$  & 4 & -1 \\
  $c$ & 4 & -1  \\
  $k_{B}$ & -5 & 1  \\
  $T$ & -5 & 1  \\ 
\end{tabular}
\label{eqn 21}
\end{equation}
The $m=2$ different $\pi$-groups are then obtained from the similarity transform $S$ (\ref{eqn 16}):
\begin{eqnarray}
	\pi_{1}&=&u^{1}\lambda^{0}h^{4}c^{4}k_{B}^{-5}T^{-5}=\frac{uh^{4}c^{4}}{k_{B}^{5}T^{5}},\nonumber\\
	\pi_{2}&=&u^{0}\lambda^{1}h^{-1}c^{-1}k_{B}^{1}T^{1}=\frac{\lambda kT}{hc}.
\label{eqn 22}
\end{eqnarray}
The inverse similarity transform $S^{-1}$ (\ref{eqn 16b}) with $x_{1}=u$ gives,
\begin{equation}
	u=\frac{k_{B}^{5}T^{5}}{h^{4}c^{4}}F(\pi_{2})=\frac{hc}{\lambda^{5}}\frac{F(\pi_{2})}{\pi_{2}^5}
\equiv\frac{hc}{\lambda^{5}}F\left( \frac{hc}{\lambda kT} \right).
\label{eqn 23}
\end{equation}
Comparing the result of DA with the exact expression for Planck's Law for the spectral energy density per unit wavelength integrated over all solid angles:
\begin{equation}
u=8\pi\frac{hc}{\lambda^{5}}\frac{1}{e(^{hc/\lambda k_{B}T})-1},
\label{eqn 24}
\end{equation}
we see that remarkable progress has been made using \textit{a priori} knowledge discovery alone. However, DA cannot provide the values of constant multipliers such as the "$8\pi$" or the actual form of the function $F\equiv 1/(e^{\pi_{2}}-1)$. And what about non-uniqueness? Since, $\pi_{2}$ is dimensionless, then so too are any algebraic combinations of $\pi_{2}$ such as $\pi_{2}^{n}$, $\frac{1}{\pi_{2}^{2}}$ or $1+\pi_{2}$ and so forth \textit{ad infinitum}. The question we are forced to ask now is the following: faced with this problem of uniqueness and the degeneracy created in the function $F$, how are we to find objective and unique dimensionless scaling relations? In particular, empirical laws published in the literature that do not resolve the problem of degeneracy should be treated with great caution. However, the information encoded into actual observational data is both sufficient and necessary to uniquely and unambiguously deduce $F$ and break the degeneracy. In a recent paper \cite{TAYLOR2007} we applied an evolutionary genetic network to a sample of 98 young galaxies where we obtained the lowest error empirical relation for their chemical abundances to date. Given a real data set, there will be one and only one best fit function $F$ which can be used to break the degeneracy inherent in the purely deductive method of DA presented in this paper.\\

\section{Conclusions}
We have generalized DA using new matrix theory that is easy to apply. We have shown how the determinant check ensures a complete specification of the problem and that the theory gives rise to a similarity transform $S$ and its inverse $S^{-1}$ that depends on a \textit{non-unique} and degenerate functional relationship $F$ between the dimensionless $\pi$ groups. The example of Planck's Law shows that DA is capable of substantial functional knowledge discovery that can be used as an intermediate step prior to data analysis, but that a full and unique solution is only possible with a correct identification of $F$ by using observational data to break its degeneracy. We hope that the simplicity of the matrix method and its ease of application will help pave the way for a new approach to the search for empirical laws and scaling relations from data.

\subsection{Acknowledgments}
MT would like to thank Jos\'e Cernicharo for facilitating this work and Pantelis Perakakis of the Universidad de Granada for useful discussions. This work was partly funded by the project "Estallidos de Formación Estelar en galaxias" (AYA2001-3939-C03) from the Spanish ministry of science.

\end{document}